\documentclass[twocolumn,aps, pra, nofootinbib,superscriptaddress,showkeys,showpacs,longbibliography]{revtex4-2}

\usepackage{textcomp}
\usepackage{setspace}
\usepackage{tipa}
\usepackage{verbatim} 
\usepackage{amsmath}
\usepackage{bbold}
\usepackage{amssymb}
\usepackage{accents}
\usepackage{amscd}
\usepackage{graphicx}
\usepackage{subcaption}
\usepackage{epstopdf}
\usepackage{caption}
\usepackage{subfig}
\usepackage{ dsfont }
\usepackage{multirow}
\usepackage{dsfont}
\usepackage{physics}
\usepackage{amsfonts}
\usepackage{graphicx}
\usepackage{xcolor}
\usepackage{physics}
\usepackage{MnSymbol}

\def\be{\begin{equation}}
	\def\ee{\end{equation}}
\def\bea{\begin{eqnarray}}
	\def\eea{\end{eqnarray}}

\def\haf{\mbox{Haf}}

\def\Tr{\mbox{Tr}}
\def\rAngle{\rrangle}

\definecolor{darkgreen}{rgb}{0,0.4,0}

\newcommand{\upb}{Paderborn University, Integrated Quantum Optics, Warburger Str. 100, 33098 Paderborn, Germany}
\newcommand{\upbphoqs}{Paderborn University, Institute for Photonic Quantum Systems (PhoQS), Warburger Str. 100, 33098 Paderborn, Germany}
\newcommand{\prague}{FNSPE, Czech Technical University in Prague, B\v{r}ehov\'{a} 7, 119 15 Praha 1, Czech Republic}

\begin{document}

\author{Magdalena Parýzková}
\affiliation{\prague}
\author{Craig S.~Hamilton}
\affiliation{\prague}
\author{Igor Jex}
\affiliation{\prague}
\author{Michael Stefszky}
\affiliation{\upb}
\affiliation{\upbphoqs}
\author{Christine Silberhorn}
\affiliation{\upb}
\affiliation{\upbphoqs}

\date{%
    \today
}

	\title{Input phase noise in Gaussian boson sampling}

	\begin{abstract}
		
        Gaussian boson sampling is an important protocol for testing the performance of photonic quantum simulators.
        As such, various noise sources have been investigated that degrade the operation of such devices. 
        In this paper we examine a situation with phase noise between different modes of the input state leading to dephasing of the system. This models the phase fluctuations, which remain even when the mean phase is controlled.
        We aim to determine whether these phase-noisy input states still form a computationally difficult problem. To do this, we use matrix product operators to model the system, a technique recently used to model boson sampling scenarios.
        Our investigation finds that the entanglement entropy grows linearly with the number of input states even for noisy input states.
        This implies that, unlike boson loss, this form of experimentally relevant noise remains difficult to simulate with tensor networks and may allow for the demonstration of quantum advantage without the need for implementing the challenging task of input-state phase stabilization.
		
	\end{abstract}
    	\maketitle
	\section{Introduction}

    Photonic quantum devices are important platforms for realizing nonuniversal quantum computers.
    Algorithms and protocols for demonstrating superior computational performance
    over classical devices, so-called quantum advantage, have been demonstrated, and new schemes continue to be developed.

	One such protocol is boson sampling, first developed by Aaronson and Arkhipov (AABS) \cite{BS_Aaronson_2011}, where bosons enter an interferometer, undergo a linear transformation, and upon exiting the device are measured in the number basis. 
	The probability to measure an output pattern depends upon the permanent of a matrix derived from the unitary transform of the interferometer. The permanent is a \#P function and, in general, is difficult for a classical computer to calculate. Thus the output statistics are computationally hard for classical machines to recreate, and there is a computational advantage for the quantum device over the classical one. This protocol has been demonstrated in several photonic experiments \cite{Broome:2013p7136, Spring:2013p7137, Tillmann:2013p10461, Bentivegnasciadv.1400255} where single photons were used.

	The AABS protocol was then expanded by some of the present authors to utilize squeezed states of light rather than single-photon states and became known as Gaussian boson sampling (GBS), which has led to several groups performing experiments related to this protocol \cite{GBS_experiment_1_Zhong_2019, GBS_experiment_Jiuzhang_1_0, GBS_experiment_Jiuzhang_2_0, GBS_experiment_Borealis}. 
	Even though these sampling problems were originally proposed as artificial problems to prove quantum advantage, other applications have been found, such as calculating vibronic spectra, predicting molecular docking configurations and various graph problems related to 
    graph isomorphisms, dense subgraphs, and finding the number of perfect matchings 
    \cite{BS_Vibronic__spectra_Huh_2015, GBS_dense_subgraphs_Arrazola_2018,GBS_graph_isomorphism_Bradler_2021}.

	Unfortunately, current quantum devices are hindered by noise, which will affect the output results and can also reduce the classical difficulty in simulating the device, thereby eliminating any quantum advantage.
	As such, there have been various ways to quantify the amount of noise that such a protocol can tolerate and still be difficult to classically simulate. 
	
	The main source of noise for photonic platforms is photon loss, and this has been studied in various works \cite{Brod_lost_2016, noisy_GBS_noise_treshold_Qi_2020, GarciaPatron2019simulatingboson}.
	It has been found that above certain levels of loss, boson sampling becomes computationally easier to sample from as the state approaches a thermal state \cite{noisy_GBS_noise_treshold_Qi_2020}.
	Recently this problem of photon loss has been studied using matrix product states (MPSs) and matrix product operators (MPOs) to arrive at similar results \cite{lossy_BS_MPO_Oh_2021, lossy_GBS_MPO_entropy_Liu_2023, lossy_GBS_MPS_algotihtm_Oh_2024}.
	In these works the complexity of these simulations is known to be related to the amount of entanglement present in the system---the more entangled the system is, the higher the computational cost.
	The authors used the entanglement entropy (EE) between bipartitions of the state as a measure of the difficulty of sampling from the state. They agreed with previous results that 
    show that photon loss can make EE grow sublinearly with respect to the system size, making the device easy for a classical computer to simulate. 
    Another source of noise in photonic systems is the partial distinguishability of photons \cite{GBS_partialdistinguishability_shi}, which leads to lower levels of interference and thus can also make the device classically simulatable.

    In this work we study the impact of phase noise in the input state of a GBS experiment. 
    Experimentally this is motivated by hybrid integrated systems, where the sources and the integrated interferometer are not on the same chip, such as the system in \cite{stefszky2025benchmarkinggaussiannongaussianinput}. Within the interferometer we can safely assume that there is no phase noise.
	Relative phase noise between the individual input modes can decohere the input state and may make the overall system easier to computationally simulate. 
	The nature of this noise source can depend upon the physical architecture of the system, for example, in spatial implementations this could be due to changes in the path lengths and in temporal architectures due to changes in the pump phase.

    Phase noise can be countered by stabilization techniques, which is a common engineering task, but when combining the lock requirements with single photon detectors and large systems, it may no longer be scalable or may lead to unwanted impacts.
    By removing phase stabilization, the state becomes susceptible to noise sources such as drifts in the phase of the pump laser or temperature drifts in the system, both leading to uncontrolled phase shifts. 
    These drifts may occur on the timescale of a single measurement run or over many hours of experimental data collection.
    Furthermore, the ability of any stabilization scheme is limited by technical details such as feedback loop bandwidth and stability requirements. Therefore, although the mean phase values can be well controlled, some uncertainty in the input phases will always remain.
    Thus, this is an experimentally relevant source of noise whose impact on boson sampling protocols has not yet been studied.

	This work aims to shed more light on the role of dephasing in a Gaussian boson sampling system. This is motivated by theoretical questions as well as practical issues related to the running of experiments.
	In Sec.~\ref{sec:gbs} we briefly introduce GBS and our model of noise. We then describe the main points of MPO that we use in Sec.~\ref{sec:mpo}, we present our results in Sec.~\ref{sec:results}, Sec.~\ref{sec:analytics} contains a theoretical argument for the scaling of entanglement in the system and in Sec.~\ref{sec:size} we comment on the possible dependence of phase noise on the system size. Finally, our conclusions are given in Sec.~\ref{sec:conc}.

	\section{GBS with phase noise}\label{sec:gbs}

\begin{figure}
		\centering
		\includegraphics[width=0.4\textwidth]{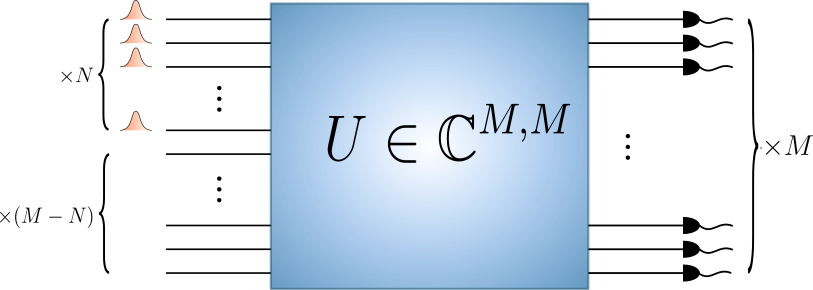}
		
		\caption{Standard Gaussian boson sampling setup, where the first $N$ input modes of an $M\times M$ interferometer are filled with Gaussian states and the other $M-N$ modes are left empty. All of the output modes are measured by photon-number-resolving detectors.}
		\label{fig:GBS_set_up}
	\end{figure}

	Gaussian boson sampling is a generalization of the standard boson sampling problem to the class of Gaussian states \cite{GBS_Hamilton_2017, GBS_detailed_study_Kruse_2019}, which includes displaced (coherent) states, thermal states, and squeezed states. 
	All such states can be described by their (Q-)covariance matrix, $\Sigma_Q$, and a displacement vector (which in this work is always zero). 
	The general experimental setup for GBS consists of an $N$-mode Gaussian state (in particular squeezed states) entering an M-mode interferometer (if $N< M$, vacuum states enter the remaining modes) and photon-number-resolving detectors at all of the output modes, see Fig.~\ref{fig:GBS_set_up}.

\begin{figure}
		\centering
		\includegraphics[width=0.35\textwidth]{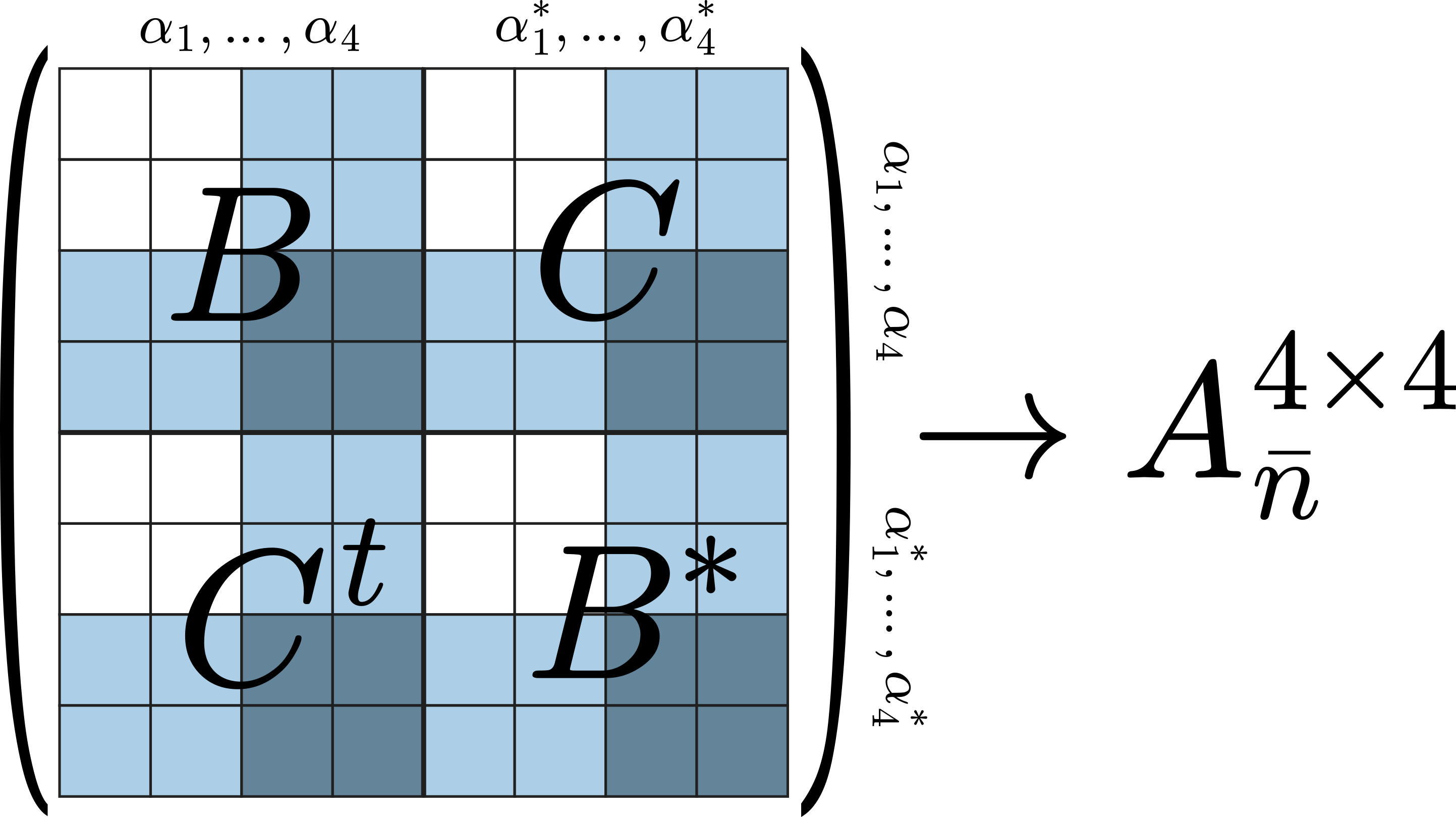}
		\caption{An example of obtaining matrix $A_{\overline{n}}$ for $M=4$ and output pattern $\overline{n} = (0,0,1,1)$. $A_{\overline{n}}$ is constructed from the dark-blue elements of the original matrix and is in this case of size $4\times 4$.}
		\label{fig:anmatrixgbs}
	\end{figure}

	The output state from the interferometer in the GBS setup is,
	\be
	\hat{\rho} = \hat{U}_{Haar}   |r\rangle \langle r|^{\otimes^N} \otimes   |0\rangle \langle 0|^{\otimes^{M-N} } \hat{U}_{Haar}^\dag,
	\ee
	where $|r\rangle \langle r| $ is the density operator of a single-mode squeezed vacuum state with squeezing parameter $r$, $|0\rangle \langle 0| $ is the single-mode vacuum state, and $\hat{U}_{Haar}$ is a unitary of the interferometer. Note that as the input state is Gaussian and the dynamics are given by a quadratic Hamiltonian, the output state remains Gaussian.
	The probability of measuring an output pattern of photon numbers, $\overline{n}=(n_1,\dots,n_M)$ (where $n_j$ represents the number of photons detected on the jth output mode), from a general quantum state can be written as, 
	\be
	\Pr(\bar{n}|\hat{\rho}) = \Tr[\hat{\rho}\hat{\bar{n}} ] = \pi^N\int d^{2N}\alpha \, Q_{\hat{\rho}}(\alpha) P_{\bar{n}} (\alpha) \label{int_overlap_QP}
	\ee

	where $Q_{\hat{\rho}}(\alpha)$ is the Q function of the state $\hat{\rho}$ and $P_{\bar{n}} (\alpha)$ is the P function of the measurement operator $\bar{n}$.
	When $\hat{\rho}$ is a Gaussian state, $Q_{\hat{\rho}}(\alpha)$ is a Gaussian function, while the $P_{\bar{n}} (\alpha)$ function of the photon number operator depends on derivatives of the Dirac $\delta$ function \cite{barnett2002methods}. When this integral is evaluated for Gaussian states, it leads to
	\begin{equation}\label{GBS_probability}
		\Pr(\overline{n}|\hat{\rho}) = \frac{1}{\overline{n}!\sqrt{\abs{\Sigma_Q}}} \text{Haf}(A_{\overline{n}}),
	\end{equation}
	where $A = \begin{pmatrix}
		0 & \mathbb{I}_M\\
		\mathbb{I}_M & 0
	\end{pmatrix}(\mathbb{I}_{2M}-\Sigma_Q^{-1})$ is a $2M\times 2M$ symmetric matrix.  
    $A_{\overline{n}}$ is constructed from $A$ in the following way. First, the rows and columns where $n_j=0$ (indices are $j$ and $j+M$) are deleted from $A$. Then the rows and columns where $n_j\ge 1$ are repeated $n_j$ times to finally obtain $A_{\overline{n}}$. An example of the whole selection process is demonstrated in Fig.~\ref{fig:anmatrixgbs}.	
	Finally, Haf($.$) denotes the Hafnian of a symmetric matrix of size $2N\times 2N$, which is a scalar function defined as
	\begin{equation}
		\text{Haf}(A) = \sum^{(2N-1)!!}_{\mu_j \in \{PMP\} } \prod^N_{k=1} A_{\mu_j(2k-1),\mu_j(2k)}.
	\end{equation}
	Here PMP denotes the set of all perfect matching permutations \cite{GBS_detailed_study_Kruse_2019}.
	Computing the Hafnian is known to be a \#P-hard problem \cite{Complexity_permanent_Valiant_1979}, and it is conjectured that even approximating the Hafnian of a matrix (up to some multiplicative error) that contains positive and negative values also belongs to the $\#$-P complexity class, making it difficult for classical computers. 
	This conjecture is based on arguments stating that if the Hafnian could be approximated easily, it would lead to the collapse of the polynomial hierarchy, which is generally considered to be extremely unlikely by the computer science community \cite{BS_Aaronson_2011, GBS_Hamilton_2017}.

    For Gaussian states the $A$ matrix is symmetric with the form,
    \begin{equation}
        A = \begin{pmatrix}
		B & C\\
	C^t & B^*
	\end{pmatrix} 
    \end{equation}
    and is of size $2M\times 2M$, where $M$ is the number of output modes measured. In the case of only squeezed light present (i.e., no thermal contribution), $C=0$ and the probability distribution is then given as,
	\begin{equation}\label{GBS_probability_B}
		P_{\hat{\rho}}(\bar{n}) = \frac{1}{\bar{n}!\sqrt{\abs{\sigma_Q}}} \abs{\text{Haf}(B_{\bar{n}})}^2. 
	\end{equation}
	In any situation with photon loss, which thermalizes the state, $C\ne 0$. 
       From this result, a GBS protocol was developed \cite{GBS_Hamilton_2017,GBS_detailed_study_Kruse_2019} that focused on sampling from squeezed vacuum states, where it was shown that a classically difficult problem occurs.
       
	\subsection{Phase noise }
	
	We now introduce our model of phase noise, which we describe as a random phase shift on each of the individual input modes. For a single mode this is
	\be
	\hat{\rho}^{(1)}_{in} = \int_\theta d\theta \, P(\theta) \, \hat{U}(\theta)\,  | r \rangle \langle r| \,\hat{U}^\dag(\theta), \label{SMSV}
	\ee
	where $| r \rangle \langle r|$ is our original single-mode squeezed vacuum state with squeezing parameter $r$, $P(\theta)$ is a probability distribution for the random phase $\theta$ and
	\be
	\hat{U}(\theta) = \exp(-i \theta \hat{n})
	\ee
    is the unitary phase-shift operator for a single mode, with $\hat{n}$ being the standard number operator. This is a bosonic dephasing channel which has been studied theoretically in other works on quantum key distribution \cite{lami2023exact}. Among other results, it has been shown that nonclassicality can still be present in such states \cite{wrapped_gaussian_distribution_Sperling_2012, kohnke_PRL_21}. 
    The level of noise can be described by the variance of $P(\theta)$, $\sigma^2$, with higher values representing noisier systems. 
    In this work we choose $P(\theta)$ to be the wrapped Gaussian distribution \cite{wrapped_gaussian_distribution_Sperling_2012}, a distribution that takes into account the finite interval over $2\pi$. However, we also considered other distributions and found that as long as the phase noise of different modes is uncorrelated, there is no appreciable difference in results.

Our GBS device is then the same as before, with $N$ of these input states, with independent noise variables $\theta_j$, entering the multimode interferometer whose unitary matrix is drawn randomly from the Haar measure. This means that the whole M-mode input state can be described as 
\begin{equation}
    \hat{\rho}_{in}= \left[ \int_\theta d^N\vec{\theta} \, P(\vec{\theta}) \, \hat{U}(\vec{\theta})\,  | r \rangle \langle r|^{\otimes N} \,\hat{U}^\dagger (\vec{\theta})\right] \otimes \ket{0}\bra{0}^{\otimes M-N},
\end{equation}
where $\vec{\theta}\equiv (\theta_1,\theta_2,\dots,\theta_N)$ and
\begin{equation}
    P(\vec{\theta})\equiv \prod_{j=1}^N P(\theta_j),\quad \hat{U}(\vec{\theta})\equiv\prod_{j=1}^N \exp (-i\theta_j \hat{n}_j)
\end{equation}
with $\hat{n}_j$ being the number operator for the jth mode.    
This leads to the state itself being a mixture of Gaussian states, which is a non-Gaussian state, and the above derivation of the photon number probability (\ref{GBS_probability_B}) does not apply directly. 
To derive a formula for $\Pr(\bar{n})$ for this state, we start from the Q function of the resulting output state $\hat{\rho}$, which is simply a linear integral,
    \begin{align}
        Q_{\hat{\rho}}(\alpha)=\int_{\vec{\theta}} d^N\theta_j \, P(\vec{\theta}) \,  Q(\alpha,\vec{\theta})\
    \end{align}
over all phase-dependent Q functions $Q(\alpha,\vec{\theta})$ of Gaussian states $\hat{\rho}(\vec{\theta})$ defined as 
\begin{equation}
    \hat{\rho}(\vec{\theta})=\hat{U}_{Haar}\left[\hat{U}(\vec{\theta})\,  | r \rangle \langle r|^{\otimes N} \,\hat{U}^\dagger (\vec{\theta}) \otimes \ket{0}\bra{0}^{\otimes M-N}\right]\hat{U}_{Haar}^\dagger.
\end{equation}
This changes (\ref{int_overlap_QP}) to
	\be
	\Pr(\bar{n}|\hat{\rho})= \int d^2\alpha\,  d^N \vec{\theta} \, P(\vec{\theta}) Q(\alpha, \vec{\theta}) P_{\bar{n}}(\alpha)
	\ee
	and leads to a formula for the probability to measure the photon number pattern $\bar{n}$:
	\be
	\Pr(\bar{n}|\hat{\rho})=  \frac{1}{\overline{n}!\sqrt{\abs{\sigma_Q}}} \int d^N \vec{\theta} \, P\left( \vec{\theta} \right) \,  \abs{\haf(B(\vec{\theta})_{\bar{n}})}^2.
	\ee
	
	We would expect this expression to be difficult to calculate for a low amount of noise, as the formula tends to the Hafnian of a single term. However, as noise increases it is not clear how to quantify the complexity of such a calculation. 
	To answer this question we use MPO to model the system and its dynamics, allowing us to calculate the MPO EE that is generated within the system. 
    If the growth of the MPO EE scales linearly (or faster) with system size, it is believed that it is not possible to simulate the system efficiently on a classical computer. On the other hand, a logarithmic scaling of MPO EE with system size strongly indicates that the system can be efficiently simulated by classical means. Photon loss in certain regimes leads to this type of scaling \cite{lossy_BS_MPO_Oh_2021, lossy_GBS_MPO_entropy_Liu_2023, lossy_GBS_MPS_algotihtm_Oh_2024}.

	\subsection{Wrapped Gaussian phase distributions}

As our model requires a probability distribution over the finite interval $[-\pi, \pi )$ and not the whole $\mathbb{R}$, the standard Gaussian distribution needs to be modified. One possibility is to use the wrapped Gaussian distribution \cite{wrapped_gaussian_distribution_Sperling_2012},
\begin{equation}\label{Wrapped_Gaussian}
			P_{WG}(\theta) = \frac{1}{\sqrt{2\pi\sigma^2}}\sum_{k=-\infty}^{\infty} e^{-\frac{(\theta-\theta_0+2\pi k)^2}{2\sigma^2}}
	\end{equation}
where $\theta_0$ is the mean value and $\sigma^2$ is the variance of the original Gaussian. The actual variance of the wrapped distribution is $1-e^{-\sigma^2/2}$. 
Also, this model implies that we know the mean value of the phase noise, which we will assume is zero here.

This mean value can usually be determined and controlled using various stabilization schemes, e.g., as in \cite{Litvin25PhaseStabilisation}.

Combined with the definition of the phase-noise channel (\ref{SMSV}), the initial state dephased under the wrapped Gaussian phase distribution for $\theta_0 = 0$ can be written as
\begin{equation}
	\hat{\rho} = \sum_{m,n=0}^{\infty}e^{-\frac{\sigma^2(m-n)^2}{2}} \rho_{mn}| m \rangle \langle n|,
\end{equation}
	where $\rho_{m,n}$ is the original density matrix element of the pure squeezed state.
    Physically, this model could represent a situation where a nonideal phase stabilization method is used, which leads to high-frequency fluctuations that remain when actively stabilizing the inputs. 

In the limit of $\sigma \rightarrow \infty$ one will get the uniform phase distribution $P(\phi) = \frac{1}{2\pi}$, which leads to the completely dephased state. Various other probability distributions were also tested but lead to similar results as the ones presented here.

	In the next section we introduce the MPO techniques that we use to calculate the EE growth in our noisy system.

	\section{MPO formalism}\label{sec:mpo}
	
	\subsection{Matrix product states and operators}
	MPSs are an important tool for simulating many-body quantum systems.
	For a pure state of a quantum many-body system with $M$ modes,
	\begin{equation}
		\ket{\psi} = \sum_{n_0,n_1,\dots,n_{M-1}} c_{n_0,n_1,\dots,n_{M-1}}\ket{n_0,n_1,\dots,n_{M-1}},
	\end{equation}
	where probability amplitude $c_{n_0,n_1,\dots,n_{M-1}}$ is an M-dimensional tensor, the MPS representation of $c$ is as follows:
	\begin{align}\label{MPS_decomposition}
		& c_{n_0,n_1,\dots,n_{M-1}}  \nonumber \\ &=\sum_{\alpha_0,\alpha_1,\dots,\alpha_{M}} \Gamma_{\alpha_0 \alpha_1}^{[1]n_0} \lambda^{[1]}_{\alpha_1}\Gamma_{\alpha_1 \alpha_2}^{[2]n_1} \lambda^{[2]}_{\alpha_2}\dots \lambda^{[M-1]}_{\alpha_{M-1}} \Gamma_{\alpha_{M-1} \alpha_M}^{[M]n_{M-1}}.
	\end{align}
	Tensors $\Gamma^{[j]}$ contain information about the jth body of the system, and vectors $\lambda^{[j]}$ are the Schmidt coefficients for the bipartition of the system at site $j$:
	\begin{equation}\label{Schmidt_decomposition}
		\ket{\psi} =  \sum_{\alpha_k}\lambda^{[j]}_{\alpha_k} \ket{\psi_{\alpha_k}^{[1,\dots,k]}}\ket{\psi_{\alpha_k}^{[k+1,\dots,M]}}.
	\end{equation}
	The maximum size of $\alpha_k$ is called the bond dimension $\chi$, and together with the dimension of the single-mode Hilbert spaces $d$ determines the complexity of MPS simulation algorithms. 
    It is important to note that in our case the tensor $\Gamma^{[1]}$ is usually only nonzero for one index $\alpha_0$, and the same holds for $\Gamma^{[M]}$ and index $\alpha_M$. The summation over indices $\alpha_0$ and $\alpha_M$ is then kept solely for generality. The memory required to store the MPS decomposition is $O(\chi^2 dM + \chi (M-1))$, and updating the MPS representation by applying some unitary operation involving at most two sites followed by single-value decomposition requires the computational time $O(d^4\chi^3)$ and $O(d^3\chi^3)$, respectively. 
	Applying an arbitrary unitary operation on the whole system can be performed by decomposing it to a series of operations involving at most two sites of the system. 
	Let $U_{k,k+1}$ be some unitary transformation on the sites $k$ and $k+1$ and $U_{n_k,n_{k+1}}^{m_k,m_{k+1}} \equiv \bra{m_k,m_{k+1}}U_{k,k+1}\ket{n_k,n_{k+1}}$. Then one first needs to calculate the tensor $\Theta$,
	\begin{align}
	   &\Theta_{\alpha_{k-1}\alpha_{k+1}}^{m_k,m_{k+1}} = \nonumber \\
       &=\sum_{n_k,n_{k+1}=0}^{d-1}\sum_{\alpha_k=0}^{\chi-1} U_{n_k,n_{k+1}}^{m_k,m_{k+1}} \lambda^{[k-1]}_{\alpha_{k-1}} \Gamma_{\alpha_{k-1} \alpha_k}^{[k]n_k} \lambda^{[k]}_{\alpha_k}\Gamma_{\alpha_k \alpha_{k+1}}^{[k+1]n_{k+1}} \lambda^{[k+1]}_{\alpha_{k+1}}, 
	\end{align}
    and then perform the singular value decomposition of $\Theta$ to recover the MPS of the evolved state (denoted by tildes),
	\begin{align}
		\Theta_{\alpha_{k-1}\alpha_{k+1}}^{m_k,m_{k+1}} &= \sum_{\beta_k=0}^{d\chi -1} V_{(m_k,\alpha_{k-1}),\beta_k}\tilde{\lambda}_{\beta_k}^{[k]} W_{\beta_k,(m_{k+1},\alpha_{k+1})},\nonumber\\
		\tilde{\Gamma}_{\alpha_{k-1} \alpha_k}^{[k]n_k} &= V_{(m_k,\alpha_{k-1}),\beta_k}\,/\,\lambda^{[k-1]}_{\alpha_{k-1}} ,\\
		\tilde{\Gamma}_{\alpha_{k} \alpha_{k+1}}^{[k+1]n_{k+1}} &= W_{\beta_k,(m_{k+1},\alpha_{k+1})}\, /\, \lambda^{[k]}_{\alpha_k} \nonumber.
	\end{align}
	As a result of this, implementing $D$ layers, each consisting of $L$ beamsplitters, on the initial state has a computational cost of $O(DLd^4\chi^3)$.
	
	Matrix product operators (MPO) are a generalization of MPS to mixed states and are necessary to simulate lossy systems. For any arbitrary state described by a density matrix $\hat{\rho}$, a similar simulation process as for pure states can be built. This can be achieved by vectorization of the density operator in the following way:
	\begin{equation}
		\hat{\rho} = \sum_{k,k'}\rho_{k,k'}\ket{k} \bra{k'} \rightarrow |\rho \rAngle = \sum_{k,k'}\rho_{k,k'}\ket{k,k'} \equiv \sum_{K} \rho_K \ket{K},
	\end{equation}
	where $K$ now denotes the double index consisting of $k$ and $k'$. Application of unitary operations $\hat{U}$ are now modified in the following way:
	\begin{equation}
		(\hat{U} \hat{\rho} \hat{U}^\dagger)_{jj'} = \sum_{kk'} U_{j,k}\rho_{k,k'}U^\dagger_{k',j'} \rightarrow (\mathcal{U}|\rho\rAngle)_J = \sum_K \mathcal{U}_{J,K}\ket{K},
	\end{equation}
	where $\mathcal{U}_{J,K} = U_{j,k}U^\dagger_{j',k'}$. The simulation then runs as described before; however, the computational complexity of applying local unitary operation now increases from $O(d^4 \chi^3)$ to $O(d^8 \chi^3)$. This is unfortunate, especially for systems with higher local dimensions $d$, such as the case discussed in this work. Fortunately, the computational cost can be reduced if the system has some global symmetry \cite{Tensor_networks_global_symmetry,Tensor_networks_U1_symmetry,Tensor_networks_SU2_symmetry}. In our case the global symmetry is the total particle number conservation which falls under the class of U(1) symmetries. 
	The complete description of how the simulation protocol can be modified in this case has been discussed in \cite{lossy_GBS_MPO_entropy_Liu_2023,lossy_BS_MPO_Oh_2021}. In short, the particle number conservation allows the operators acting on the systems to be written as a direct sum $\hat{T} = \bigoplus_n \hat{T}_n$, where $\hat{T}_n$ acts on the subspace of the total Hilbert space corresponding to $n$ particles in the system. This means that all tensors can be written in block-diagonal form. Consequently, they can be stored and manipulated more efficiently.

	\begin{figure}
		\centering
		\includegraphics[width=0.5\textwidth]{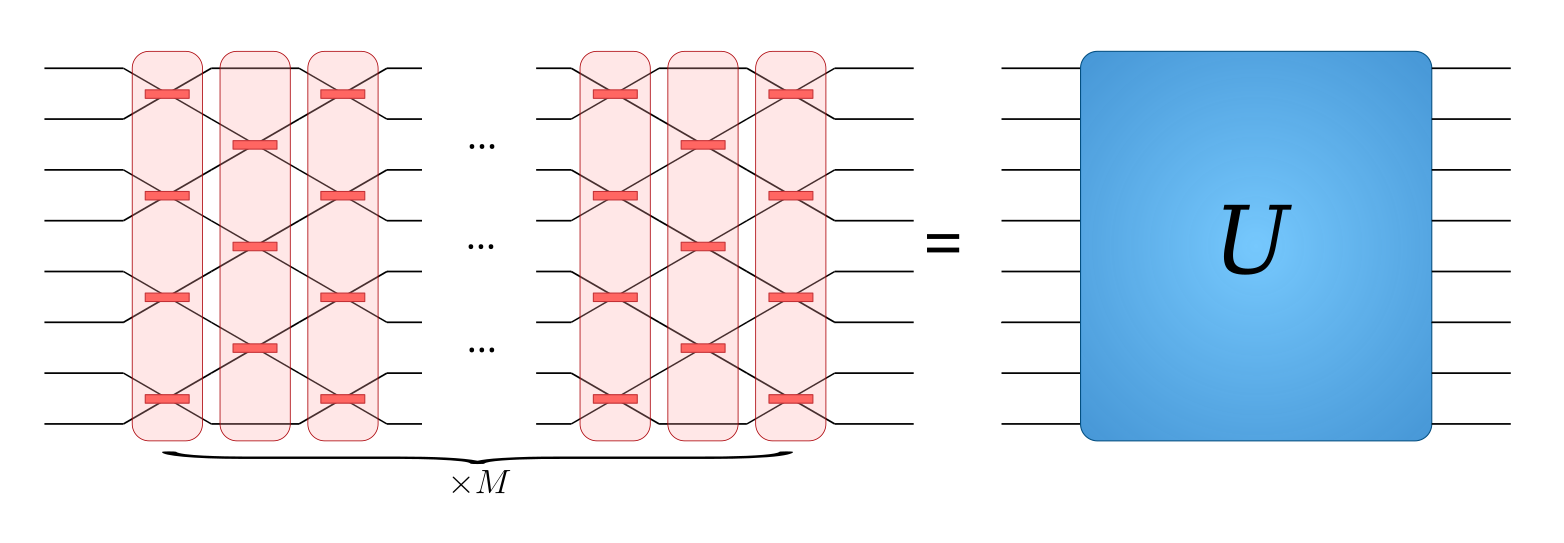}
		\caption{Haar random interferometer of size $M\times M$ corresponding to unitary operation $U$ decomposed into $M$ layers of beamsplitters, which in this case represent local operations on two modes.}
		\label{fig:beamsplittersinterferometer}
	\end{figure}

	\subsection{Operator entanglement entropy}
	The efficiency of MPS simulations for a given family of states is determined mainly by asymptotic scaling of the bond dimension, $\chi$, with the systems size. If $\chi$ grows exponentially with the size of the system, then the MPS simulation becomes inefficient. 
	It has been previously established that the growth of the bond dimension in simulating many-body quantum systems is closely related to the R\'enyi entropy $S_\alpha$ of different bipartitions of the system \cite{Entropy_scaling_and_MPS_Schuch_2008,MPO_Approximability_Jarkovsky_2020,MPS_Approximability_Vestraete_2006}, where the R\'enyi entropy of a quantum state $\hat{\rho}$ is defined as
	\begin{equation}\label{Renyi_EE_state}
		S_\alpha(\hat{\rho}) = \frac{1}{1-\alpha} \log_2 \left( \Tr \hat{\rho}^\alpha\right), \quad \alpha\in\left( 0,1 \right) \cup \left(1,\infty \right),
	\end{equation}
	and in the limit $\alpha\rightarrow 1$ it tends to the von Neumann entropy,
	\begin{equation}\label{von_Neumann_EE_state}
		S_1(\hat{\rho}) \equiv S (\hat{\rho}) = - \Tr(\hat{\rho} \log_2 \hat{\rho}).
	\end{equation}
	More specifically, for pure states the R\'enyi entropy quantifies the amount of entanglement present in the system. Thus, for more entangled states a higher bond dimension is necessary to represent the system with MPS.
	The details of the relationship between the asymptotic scaling of R\'enyi entropies with system size and the efficiency of MPS simulations has been investigated in \cite{Entropy_scaling_and_MPS_Schuch_2008,MPS_Approximability_Vestraete_2006}. 
    For example, for von Neumann entropies, which will be studied here, linear (and faster) growth means that the system is difficult to simulate using MPS. On the other hand, logarithmic (and slower) asymptotic behavior for $\alpha<1$ means that the system can be efficiently simulated with MPS.

	Here we will be interested in the R\'enyi (and also von Neumann) entropy, maximized over all bipartitions of the states, throughout the whole time evolution, defined as, 
    	\begin{equation}\label{EE}
		S_\alpha^{max}(\ket{\psi}) = \max_{1\leq k<M}S_\alpha (\Tr_{1,\dots,k}[\ket{\psi}\bra{\psi}]).
	\end{equation}

As mentioned earlier, the MPS decomposition (\ref{MPS_decomposition}) is especially convenient for calculating the R\'enyi (and von Neumann) entropies for different bipartitions, since $\lambda^{[j]}$ are the vectors of Schmidt coefficients for different bipartitions of the system (see Eq.~(\ref{Schmidt_decomposition}) for details). 
	Using the Schmidt coefficients, the R\'enyi entropy of a pure state $\rho$ can be calculated as,
	\begin{equation}\label{Renyi_EE_lambda}
		S_\alpha(\hat{\rho}) = \frac{1}{1-\alpha} \log_2 \left( \sum_k \lambda_k^\alpha\right)
	\end{equation}
	and similarly for the von Neumann entropy,
	\begin{equation}\label{von_Neumann_EE_lambda}
		S(\hat{\rho}) = -\sum_k \lambda_k \log_2 \lambda_k.
	\end{equation}

     As was discussed in \cite{lossy_BS_MPO_Oh_2021,lossy_GBS_MPO_entropy_Liu_2023,MPO_Approximability_Jarkovsky_2020}, for MPO simulations the situation becomes more complicated.
    When considering MPO simulations the tensors $\lambda^{[j]}$ represent the following decomposition:
	\begin{equation}
		\hat{\rho} =  \sum_{\alpha_k}\lambda^{[k]}_{\alpha_k} \hat{\rho}_{\alpha_k}^{[1,\dots,k]}\hat{\rho}_{\alpha_k}^{[k+1,\dots,M]}.
	\end{equation}
	Due to the vectorization of the density matrix, the sum $\sum_{\alpha_k}\abs{\lambda^{[k]}_{\alpha_k}}^2$ is constant during the MPO simulation but does not necessarily need to be equal to 1 anymore; therefore before calculating entropies we normalize each vector $\lambda^{[k]}$ to 1.
	From these $\lambda$ coefficients one can again calculate the R\'enyi or von Neumann entropy of the kth bipartition defined by formulas (\ref{von_Neumann_EE_lambda}) and (\ref{Renyi_EE_lambda}). However, in the MPO case they are not the same as pure state entropies (\ref{Renyi_EE_state}) and (\ref{von_Neumann_EE_state}). Instead, they are known in the literature as operator entanglement entropies \cite{dephasing_MPO_22,operator_entanglement_entropy_equilibrium,operator_entanglement_entropy_chaos_Zhou_2017,operator_entanglement_entropy_chaos_Wang_2019}.
	In the case of pure states the operator entanglement entropy is twice the state entanglement entropy, but for mixed states they are two different quantities.
	Also, unlike for pure states, mixed-state operator entanglement entropies are not necessarily connected to genuine quantum entanglement, since they can be influenced by classical correlations as well. However, they can still give insights into quantum many-body effects, such as quantum chaos and
	information scrambling \cite{operator_entanglement_entropy_chaos_Zhou_2017, operator_entanglement_entropy_chaos_Wang_2019, operator_entanglement_entropy_equilibrium}. 

    As a consequence of this difference, the results of \cite{Entropy_scaling_and_MPS_Schuch_2008} cannot be applied directly. Due to vectorization of the density matrix, operator entanglement entropies bound the accuracy of the approximation in terms of the vector 2-norm instead of the trace distance. The relationship between these two norms, $\left\Vert\hat{\rho}\right\Vert_1 \leq d^{M/2} \left\Vert\hat{\rho}\right\Vert_2$, allows a situation where a larger bond dimension is required to accurately approximate the state in terms of trace distance, while the operator entanglement entropy decreases. However, since it also holds that $\left\Vert\hat{\rho}\right\Vert_2 \leq \left\Vert\hat{\rho}\right\Vert_1$, then if the 2-norm cannot be made arbitrarily small using a polynomially scaling bond dimension, the same will be true for the trace distance. For this reason only the inefficiency of MPO simulations for linear scaling of the operator Rényi entropy of $\alpha > 1$ and the operator von Neumann entropy is guaranteed. For a detailed explanation, we refer the reader to \cite{lossy_BS_MPO_Oh_2021,MPO_Approximability_Jarkovsky_2020}.
    
    In our MPO simulations we will be calculating the operator von Neumann entropy, i.e., the value of (\ref{EE}) for $\alpha\rightarrow 1$ and where $\ket{\psi}$ is the vectorized density matrix of our system. From now on we will refer to this entropy as the MPO entanglement entropy (MPO EE).

	\section{Results: MPO simulations of phase-noisy GBS system}\label{sec:results}
		
	In this section, we present our results for the scaling of MPO EE in GBS with phase noise, using both numerical calculations with MPOs and an analytical reasoning for the behavior observed.

	We model $N$ single-mode squeezed states, with squeezing parameter $r$, and send each of them through a single-mode dephasing channel described by a probability distribution $P(\theta)$, i.e., Eq.~(\ref{SMSV}).
	The states then enter a random interferometer (of size $M\times M, M> N$), whose unitary matrix is drawn from the Haar measure. The unitary is decomposed into a series of layers, where each layer consists of $2\times 2$ beamsplitters between neighboring modes that recreate the whole network. 
	From the final output state of this network, the MPO EE between each bipartition of the $M$ modes is calculated and the maximum value is taken as characterizing the system.
    We have adapted the code and algorithm from \cite{lossy_GBS_MPO_entropy_Liu_2023} and modified it for our purposes to calculate the MPO EE of the GBS system with phase noise \cite{codeFNSPE}.

	The behavior of the MPO EE, given by (\ref{EE}), for boson sampling with photon loss present has been studied recently both for standard BS \cite{BS_MPS_Huang_2019, lossy_BS_MPO_Oh_2021} and GBS \cite{lossy_GBS_MPO_entropy_Liu_2023, lossy_GBS_MPS_algotihtm_Oh_2024}, showing in all cases photon loss influences the scaling behavior of MPO EE significantly even for small system sizes.
	The values of the local dimension $d$ and bond dimension $\chi$ for each simulation were chosen to keep the error of the total probability, $1-\Tr[\hat{\rho}]$, smaller than $1\%$, a previously used standard \cite{lossy_BS_MPO_Oh_2021}. The interferometer is simulated as a Haar random unitary in $M$ layers of beamsplitters according to \cite{Haar_random_in_layers_Russell_2017}, see Fig.~(\ref{fig:beamsplittersinterferometer}). There they demonstrate a way to directly generate the layers of beamsplitters, so it is not necessary to generate the complete Haar random unitary and then decompose it. 
	Following this for a single run, the results are then averaged over multiple runs. In all of the simulations the maximum MPO EE from all layers and over all possible bipartitions is found, as it represents the maximum bond dimension reached while simulating the full system.

\begin{figure}
		\centering
		\includegraphics[width=0.5\textwidth]{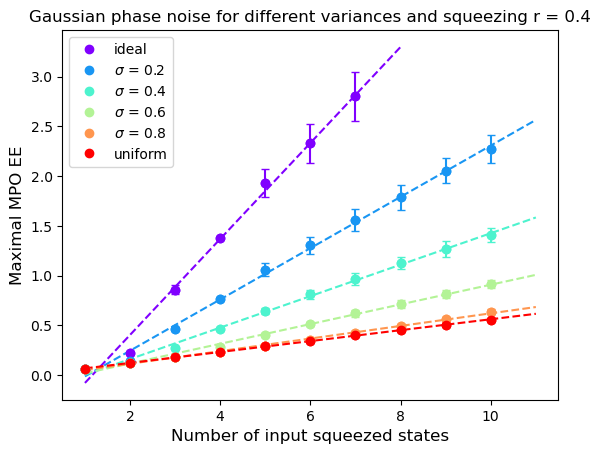}
		\caption{Simulation of maximal MPO EE for phase-noisy input squeezed states with fixed squeezing parameter $r=0.4$ and different amounts of phase noise. The model of phase noise is a wrapped Gaussian distribution with different $\sigma$. The lines with corresponding colors are linear fits of the data.}
		\label{fig:diff_phase_noise}
	\end{figure}

	In Fig.~\ref{fig:diff_phase_noise} we plot the maximal MPO EE vs. $N$, the number of input states, for constant squeezing $r=0.4$ and with different variances $\sigma$ of the wrapped Gaussian distribution (\ref{Wrapped_Gaussian}). The size of the interferometer in the simulations is $M\times M$, where $M=\max\lbrace20,4 N\rbrace$. The squeezing parameter has been chosen because of the suitability to our current computational resources. 
	It can be seen that while the growth of MPO EE slows with higher levels of noise, the overall scaling still remains linear for a small number ($N=10$) of input states, meaning that this effect can be verified for small systems. 
	This is significantly different when compared to the results of \cite{lossy_GBS_MPO_entropy_Liu_2023}, as for photon loss the sublinear scaling of MPO EE is already visible for low numbers of input states.

\begin{figure}
		\centering
		\includegraphics[width=0.48\textwidth]{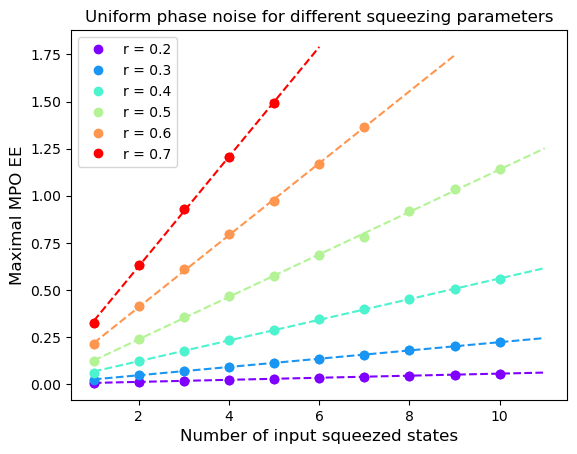}
		\caption{Simulations of maximal MPO EE for completely phase-noisy input squeezed states with different squeezing parameters $r$ (uniform distribution). The lines with corresponding colors are linear fits of the data.}
		\label{fig:completely_dephased}
	\end{figure}

	Next, we simulated the system with a maximal amount of phase noise, represented by the uniform distribution $P_U(\theta)=\frac{1}{2\pi}$. This leads to the completely dephased squeezed state, with a density matrix diagonal in the number basis. 
	The results of these simulations for different squeezing parameters $r$ can be seen in Fig.~\ref{fig:completely_dephased}. As expected, the maximal MPO EE in the system grows with the amount of squeezing. However, the scaling in the number of input modes still behaves linearly. 
    It is important to note here that in all of these simulations we are still operating in the collision-free regime required for the classical complexity of the GBS problem \cite{GBS_Hamilton_2017,GBS_detailed_study_Kruse_2019}. If the squeezing continued to increase without any further adjustments to the system size, it would eventually break this collision-free assumption and MPO EE might behave differently.

\begin{figure}
		\centering
		\includegraphics[width=1\linewidth]{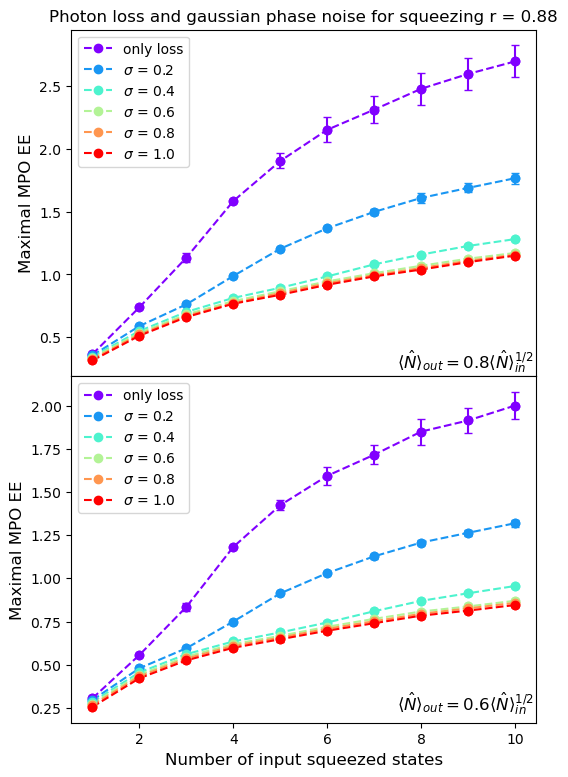}
		\caption{Simulations of maximal MPO EE for the combination of photon loss and Gaussian dephasing. On the left the photon loss scaling $\langle \hat{N} \rangle_{out} = 0.6 \langle \hat{N}\rangle_{in}^{1/2}$ is used and on the right the scaling is $\langle \hat{N}\rangle_{out} = 0.8 \langle \hat{N}\rangle_{in}^{1/2}$. 
        }
		\label{fig:lossphasenoise}
	\end{figure}

	We were unable to simulate higher levels of squeezing, e.g., as in Ref.~\cite{lossy_GBS_MPO_entropy_Liu_2023}. The reason for this is that the amount of squeezing in the system affects the necessary local dimension $d$ of the single-mode Hilbert spaces, and larger squeezing requires larger $d$ to keep the approximation of the system precise enough. While photon loss naturally decreases the local dimension, phase noise does not because it affects only the off-diagonal terms of the density matrix. 
	Finally, the variance of the MPO EE results over multiple simulations with different Haar random unitaries is low and all of the graphs are averaged over at least 100 simulations, as the resulting MPO EE has small variance. It is worth noting that the variance also diminishes noticeably with the amount of phase noise present in the system. This is demonstrated by the error bars ($\pm1$ SD) in Figs.~\ref{fig:diff_phase_noise}, \ref{fig:lossphasenoise} and \ref{fig:changingsigma}. In Fig.~\ref{fig:completely_dephased} the variance of the MPO EE is so small that the error bars are not visible.
	
	Finally, we look at how the combined simulations of both phase noise and photon loss behave. As can be seen in Fig.~\ref{fig:lossphasenoise}, the combination of photon loss and Gaussian phase noise seems to behave as expected, with dephasing slowing the growth of MPO EE and the sublinear growth remaining.  

	\section{Analytical results for MPO EE scaling for phase-noisy input states}\label{sec:analytics}
	
	In Refs.~\cite{lossy_BS_MPO_Oh_2021, lossy_GBS_MPO_entropy_Liu_2023} the authors studied the analytical asymptotic scaling of MPO EE for AABS and GBS. For GBS with uniform loss present, the asymptotic scaling was shown to be
	\begin{align}\label{EE_scaling_loss}
		S_1(\hat{\rho}_{out}) &=\begin{cases}
			O\left( N^{2\gamma-1}\log_2(N)\right)  & \dots 0<\gamma\leq 1, \\
			O(N) & \dots \gamma = 1,
		\end{cases} \nonumber\\
		S_\alpha(\hat{\rho}_{out}) &= \begin{cases}
			O\left( N^{1-2(1-\gamma)\alpha}\right)  & \dots \alpha<1, \\
			O\left( \frac{\alpha}{1-\alpha}N^{2\gamma-1}\right)  & \dots \alpha>1,
		\end{cases}
	\end{align}
	which corresponds to their numerical results, and where $N$ is the number of input squeezed states and $\langle \hat{N}\rangle_{out}  = \beta \langle \hat{N}\rangle_{in}^{\gamma}$ being the overall photon loss scaling. Here $\langle \hat{N}\rangle_{in}$ and $\langle \hat{N}\rangle_{out}$ are the average total number of photons in the system before and after the interferometer.
	
	We follow the general procedure found in \cite{lossy_BS_MPO_Oh_2021, lossy_GBS_MPO_entropy_Liu_2023, park2025matrix} to analytically determine the MPO EE scaling for the collision-free regime \cite{BS_Aaronson_2011}.
	We assume that our input state is composed of $N$ single-mode states 
	\begin{equation}
		\hat{\rho}_{in} = \otimes^N_{j=1} \hat{\rho}_{j,in}, \quad \text{where } \hat{\rho}_{j,in} = \sum_{n,m=0}^{\infty}C_{nm}\ket{n}_j\bra{m}_j.
	\end{equation}
	and we first apply our noise channel $\epsilon_{noise}$ (photon loss, dephasing, etc.) to this state
	\begin{equation}
		\hat{\rho}_{in,noise} = \epsilon_{noise} (\hat{\rho}_{in}).
	\end{equation}

	Next, the M-mode interferometer, represented by unitary $\hat{U}$, is applied. Its effect on the system can be described by the transformation of the input creation operators $\hat{a}_k^{\dagger}$ into the creation operators $\hat{b}_k^{\dagger}$ of the output modes as $\hat{b}_j^{\dagger} = \sum_{k=1}^{M} U_{jk} \hat{a}_k^{\dagger}$.
	To study the specific bipartition of the system at the $l$th site after the interferometer is applied, one can define new ``up'' and ``down'' creation operators $ \hat{B}^{\dagger}_{u,j} $ and $ \hat{B}^{\dagger}_{d,j} $
	\begin{equation}
		\cos \theta_j \hat{B}^{\dagger}_{u,j}=\sum_{k=1}^{l} U^{-1}_{jk} \hat{b}_k^{\dagger}, \quad \sin \theta_j \hat{B}^{\dagger}_{d,j}=\sum_{k=l+1}^{M} U^{-1}_{jk} \hat{b}_k^{\dagger},
	\end{equation}
	with 
	\begin{equation}
		\cos^2 \theta_j = \sum_{k=1}^{l} \abs{U^{-1}_{jk}}^2, \quad \sin^2 \theta_j = \sum_{k=l+1}^{M} \abs{U^{-1}_{jk}}^2.
	\end{equation}
	For the collision-free case, when one measures at most one photon in each mode at the output, the following canonical commutation relations hold for large M
	\begin{align}
		\left[ \hat{B}_{u,j}, \hat{B}^{\dagger}_{u,k}\right]  \approx \delta_{jk},\quad\left[ \hat{B}_{d,j}, \hat{B}^{\dagger}_{d,k}\right]\approx \delta_{jk}, \nonumber\\
        \quad \left[ \hat{B}_{u,j}, \hat{B}_{d,k}\right]  = \left[ \hat{B}_{u,j}, \hat{B}^{\dagger}_{d,k}\right] =0,
	\end{align}
	so one can define the approximately orthogonal bipartition number states
	\begin{equation}
		\frac{\hat{B}^{\dagger k}_{side,j}}{\sqrt{k!}}\ket{0}_j = \ket{k}_{side,j},\quad side\in \left\lbrace u,d \right\rbrace,
	\end{equation}
	where $\ket{k}_{side,j}$ now represents the state of $k$ photons originally from the jth mode on one particular side of the bipartition. The standard Fock states of the jth mode are transformed as follows
	\begin{equation}
		\ket{n}_j \rightarrow \sum_{k = 0}^{n} \sqrt{\binom{n}{k}}\cos^{k}\theta_j \sin^{n-k}\theta_j \ket{k}_{u,j}\ket{n-k}_{d,j}.
	\end{equation}

    For an arbitrary separable input state composed of M single-mode input states one can write
    \begin{align}
        &\ket{\psi_{in}}= \otimes_{j=1}^{M}\sum_{n_j=0}^{\infty}A_{n_j}\frac{(\hat{a}^\dagger_j)^{n_j}}{\sqrt{n_j!}}\ket{0} \nonumber \\ 
        &\rightarrow \ket{\psi_{out}} = 
        \otimes_{j=1}^{M}\sum_{n_j=0}^{\infty}A_{n_j}\frac{(\cos^2 \theta_j \hat{B}^{\dagger}_{u,j}+\sin^2 \theta_j \hat{B}^{\dagger}_{d,j})^{n_j}}{\sqrt{n_j!}}\ket{0} \nonumber\\
        &=\otimes_{j=1}^{M}\sum_{n_j=0}^{\infty}A_{n_j} \sum_{k = 0}^{n_j} \sqrt{\binom{n_j}{k}}\cos^{k}\theta_j \sin^{n_j-k}\theta_j \ket{k}_{u,j}\ket{n_j-k}_{d,j},
    \end{align}
    where $A_{n_j}$ are probability amplitudes for individual single-mode input states.
Therefore, if the noise channel preserves the separability of the initial state, the output state then will be separable in the new ``up'' and ``down'' formalism as well
    \begin{equation}
    \hat{\rho}_{out} = \otimes_{j=1}^N\hat{\rho}_{j,out}, \quad \text{where   } \hat{\rho}_{j,out}=\hat{U}\epsilon_{noise}(\hat{\rho}_{j,in})\hat{U}^{\dagger}.
    \end{equation}
    Consequently, the R\'enyi entropy (\ref{EE}) can be calculated as a sum of entropies originating from the individual input modes
	\begin{equation}
		S_\alpha(\hat{\rho}_{out}) = S_\alpha(\otimes^N_{j=1}\hat{\rho}_{j,out}) = \sum^N_{j=1}	S_\alpha(\hat{\rho}_{j,out}).
	\end{equation}
	This means that it is sufficient to calculate the scaling of single-mode entropies in the new basis. Note that in our case, where we seek to obtain the scaling of MPO EE, the exact calculation includes vectorizing the density matrix in addition to tracing out one part of the bipartition. 
	
	That being said, one can notice that even for identical input states, the outputs $\hat{\rho}_{j,out}$ will no longer be identical because of their dependence on the specific choice of $\hat{U}$. However, since $\hat{U}$ is an arbitrary random unitary independent of $N$, the scaling remains the same for all $\hat{\rho}_{j,out}$. For large $N$ this allows us to approximate
	\begin{equation}
		S_\alpha(\hat{\rho}_{out}) \approx N S_\alpha(\hat{\rho}_{1,out}),
	\end{equation}
    where $\hat{\rho}_{1,out}$ was used to denote common asymptotic scaling of all $\hat{\rho}_{j,out}$.

	In Refs.~\cite{lossy_BS_MPO_Oh_2021, lossy_GBS_MPO_entropy_Liu_2023} the photon loss channel creates a dependency of $\hat{\rho}_{j,out}$ on the total number of input modes $N$, which then results in potential sublinear scaling of MPO EE (\ref{EE_scaling_loss}). However, our phase-noise channel (\ref{SMSV}) does not depend on the total number of input modes at all. Thus the elements of $\hat{\rho}_{j,out}$ do not depend on $N$ as well and consequently have a constant scaling.
    This means that the scaling of MPO EE will always be linear for this particular model of dephasing, independent of the probability distribution $P(\theta)$. In fact, any noise channel which affects input states locally mode-by-mode and in a way which is independent of the number of modes present in the system will keep its original scaling. Consequently, this model of phase noise makes the GBS system difficult to simulate with tensor networks. 

	We can use the methods of this section to calculate the specific asymptotes for a given squeezing parameter, dephasing model and a number of input modes. Higher initial dephasing in the input causes the entropy contribution from a single mode to be lower, and so the linear scaling of total entropy (\ref{EE_scaling_loss}) is less steep. This is in agreement with our numerical results, in Sec.~\ref{sec:results}.
 
    We have made preliminary investigations into the speed of convergence of our MPO simulation results to the specific asymptotics calculated using the results from this section similarly as it was done in \cite{lossy_GBS_MPO_entropy_Liu_2023}. While for photon loss the convergence has been shown to be quite fast \cite{lossy_GBS_MPO_entropy_Liu_2023}, for phase noise it seems to be much longer. This is likely caused by the fact that photon loss by its definition inherently destroys the probability of measuring multiple photons in any of the output modes, making the system converge faster to the collision-free asymptotics, while phase noise does not. Combined with the exponential increase of computational cost of the simulations with the system's size and our limited computational resources, the numerical confirmation of convergence unfortunately remains beyond the scope of this paper at the moment. However, study of the exact mechanism of convergence for the phase-noise case is of interest for future investigations.

It is also important to mention that any of the arguments regarding the complexity of MPO simulations do not take into consideration the scaling of the local dimension $d$, since entanglement entropy only reflects the scaling of bond dimension $\chi$. 
    
\section{Dependence of phase noise on the system size}\label{sec:size}
     Until now we have considered the phase-noise model to be independent of the system size, i.e., the number of input modes.
    One may expect that, as for photon loss, the dephasing may increase as we scale up the system. 
    In this section we will explain why this does not influence our result of linear asymptotic scaling of MPO EE and consequently the inefficiency of MPO simulations for Gaussian boson sampling with dephased input states.

    For any increasing amount of phase noise, the phase distribution will tend to the uniform one. Thus we can say that this represents the worst-case scenario for this noise model. 
    As can be seen in Fig.~\ref{fig:diff_phase_noise}, with an increasing amount of phase noise the dephasing model (\ref{SMSV}) will always converge to the uniform phase distribution $P(\phi) = 1/2\pi$. 
    Consequently, the MPO EE resulting from the dephasing model (\ref{SMSV}) with arbitrary phase distribution $P(\phi)$ will always be larger, or equal to, the MPO EE resulting from the phase noise of the uniform phase distribution. This means that even though the MPO EE scaling can (and probably will) have sublinear behavior in the beginning, in the limit $N\rightarrow \infty$ it will converge to the linear asymptotic scaling of the uniform phase distribution. 

    \begin{figure}
        \centering
        \includegraphics[width=1\linewidth]{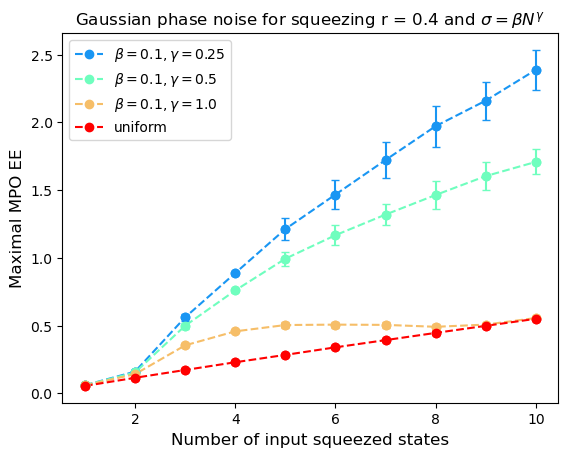}
        \caption{Simulations of maximal MPO EE for the phase noise dependent on the number of input states $N$. The model of dephasing is the wrapped Gaussian distribution (\ref{Wrapped_Gaussian}) with variance monotonously increasing as $\sigma = \beta N^{\gamma}$.}
        \label{fig:changingsigma}
    \end{figure}

    We demonstrate this argument on the model of the wrapped Gaussian distribution, with $\sigma = \sigma(N)$ being a monotonously increasing function of $N$, e.g., $\sigma = \beta N^{\gamma}$ for $\beta, \gamma > 0$. In Fig.~\ref{fig:changingsigma} one can see simulations for the choice of $\beta = 0.1$ and $\gamma \in \lbrace 1/4, 1/2, 1\rbrace$. One can clearly see that at the beginning MPO EE drops to sublinear scaling as expected. However, for $\gamma = 1$ one can already see convergence to the uniform phase distribution even for small $N$. The size of the interferometer in all simulations is again $M=\max\lbrace20,4 N\rbrace$, and the resulting MPO EE has been averaged over at least 100 Haar random interferometers.

	\section{Summary and Conclusions}\label{sec:conc}

	In this work we have analyzed the problem of relative phase noise within an input state of a Gaussian boson sampling protocol. 
	If the phase of the pump laser is not well defined or the path lengths of the time-space demultiplexing device, then phase drift and fluctuations can occur between the different modes of the state, a mechanism which decoheres the initial state before it enters the interferometer. This may lead to the quantum system being classically simulatable, as is the case for photon loss, which changes the scaling of MPO EE from linear to sublinear.

	Here we have shown that the initial phase noise leads to the MPO EE growing linearly, which will still lead to a problem that is classically difficult to sample from using tensor networks. We confirmed this result both numerically and analytically.
	One possible explanation is that in the limit of complete dephasing 
	the state is still nonclassical with entanglement present, whereas in the limit of large photon loss the squeezed thermal state can have no entanglement present.

Our work shows that some sources of noise may not lead to a classically simulatable problem in the limit of large noise. 
    We have shown that dephasing does not change the scaling of MPO EE growth. 
    This suggests that phase stabilization is not a critical factor when designing experimental systems.
    This may reduce the requirements of the phase control scheme implemented on the input states in the context of classical complexity---although one needs to control the mean of the phase, the system is largely insensitive to fluctuations about this mean.

	In future work we will study how photon number correlations between modes can differentiate between levels of dephasing in an experiment and also distinguish between quantum and classical states. We can also extend the model to investigate the 
     role of continuous dephasing throughout an experimental setup. 
     This is important work for the future development of GBS using minimal experimental resources, as it allows us to reduce the requirements of the feedback systems used to control these devices.
	
	\section{Acknowledgments}
     I.J., C.S.H., C.S., and M.P. acknowledge the financial support by the Horizon-CL4 program under Grant Agreement No. 101135288 for the EPIQUE project, and from the state budget of the Czech Republic under RVO14000. I.J. and M.P. acknowledge support of the Grant Agency of the Czech republic (GA\v{C}R) under Grant No. 23-07169S.

    This work has received funding from the German Federal Ministry of Research, Technology and Space (BMFTR) within the PhoQuant project (Grant No. 13N16103).

    The authors would also like to thank Minzhao Liu and Changhun Oh for making their code available and for valuable discussions. 

    \section*{Data Availability}
    The code used to generate data which support the findings of this article is openly available \cite{codeFNSPE}.

	\bibliographystyle{IEEEtran}
	\bibliography{bibliography}
	
\end{document}